\documentclass[prb,twocolumn,showpacs]{revtex4}
\usepackage{graphicx}
\usepackage{epstopdf}
\usepackage{amsfonts}
\usepackage{amsmath}
\usepackage{amssymb}
\usepackage{bm}
\usepackage{hyperref}
\begin{document}
\title{Type-II Dirac cone and  Dirac cone protected by nonsymmorphic symmetry \\
  in  carbon-lithium compound (C$_{4}$Li)}
\author{Armindo S. Cuamba$^{1}$}
\author{Pavan Hosur$^{1}$}
\author{Hong-Yan Lu$^{1,2}$}
\author{Lei Hao$^{1,3}$}
\author{C. S. Ting$^{1}$}

\affiliation{$^1$ Texas Center for Superconductivity and Department of Physics, University of Houston,
 Houston, Texas 77204, USA\\
 $^2$ School of Physics and Electronic Information, Huaibei Normal University, Huaibei 235000,China\\
 $^3$ Department of Physics, Southeast University, Nanjing 210096, China
}
\date{\today }

\begin{abstract}
  In this work, we predict a novel band structure for Carbon-Lithium(C$_{4}$Li) compound  using the first-principles method.  We show that it exhibits  two Dirac points near the Fermi level; one located at W point originating from the nonsymmophic symmetry of the compound, and the other one behaves like a  type-II Dirac cone with higher anisotropy along the $\Gamma$ to X line. The obtained Fermi surface sheets of the hole-pocket and the electron-pocket near the type-II Dirac cone are separated from each other, and they would touch each other when the Fermi level is doped to cross the type-II Dirac cone. The evolution of Fermi surface with doping  is also discussed.  The bands crossing  from T to W make a  line-node at the intersection  of k$_{x}$=$\pi$ and k$_{y}$=$\pi$ mirror planes.   The C$_{4}$Li is a  novel  material with  both nonsymmorphic protected Dirac cone  and  type-II Dirac cone near the Fermi level which may exhibit exceptional topological property for electronic applications.

  \end{abstract}

  \maketitle

  \emph{Introduction}:  Carbon has been extensively investigated due to its exceptional property and easier syntheses in the related industry.  Graphite\cite{Geim-2007} and diamond are some common forms of carbon found in nature. For instance, diamond is an insulator with strong elastic properties and has been used to achieve high-pressure environment\cite{Paul-2012}. Other carbon structures were predicted to be stable\cite{Malko-2012} and synthesized under  different conditions. Graphene forms a Dirac cone at high symmetry point with zero density of states at the Fermi level. Intercalated graphene with alkaline metals such as lithium was reported to exhibit a metallic band structure and  superconductivity\cite{Gianni-2012}. In addition, carbon-lithium structure was used to generate lithium battery\cite{Hassoun-2014}.

  The realization of carbon structure with nontrivial topology has become an interesting topic recently, and different structures were  proposed. For example, the two-dimensional carbon structure; the $\alpha$-graphyne, $\beta$-graphyne, and (6,6,12)-graphyne exhibit Dirac cones\cite{Malko-2012}. The three-dimensional carbon structure; the bco-C$_{16}$\cite{Wang-2016}, and bct-C$_{16}$\cite{Cheng-2017} were predicted  to be a line-node semimetal protected by the mirror symmetry, and the graphene has type-I Dirac cone which is similar to the Dirac cone of Na$_{3}$Bi\cite{Liu-2014-Na3Bi} and Cd$_{3}$As$_{2}$\cite{Liu-2014-Cd3As3} material. The nonsymmorphic\cite{Parameswaran-2013} material was predicted to exhibit a different type  of Dirac cone  characterized by having a crossing of two-fold degenerate bands and result in a four-fold point node on HfI$_{3}$ compound\cite{Gibson}. This type of crossing is located at the high symmetry point in the Brillouin zone (BZ).  Later the Angle-resolved photoemission experiments (ARPES)   confirmed the existence of this type of Dirac cone in ZrSiS compound\cite{Schoop}. Besides, there exist other materials that are Dirac node-semimetals protected by nonsymmorphic symmetry\cite{chen-2017}.

  The Weyl semimetal material are classified as type-I\cite{Yang-2015,Xu-2016-TaP,Sun-2015} and type-II\cite{Deng-2016,Alexey-2015} crossings and both were realized. Similarly, the Dirac semimetal  can also  be classified as type-I and type-II. While many structures have been classified as  type-I Dirac semimetals\cite{Liu-2014-Cd3As3,Liu-2014-Na3Bi},  the search for type-II Dirac semimetals is still an interesting and emergent issue. Although some compounds were predicted to have type-II Dirac fermions\cite{Chang-2017,Guo-2017,Le-2016,Huang-2016}, and its existence has only been observed in PtSe$_{2}$\cite{Zhang-2017}, PtTe$_{2}$\cite{Yan-2017} and PdTe$_{2}$\cite{Han-2017}. Unfortunately, the observed Dirac cone\cite{Zhang-2017,Yan-2017,Han-2017} is located away from  the Fermi level that would make it be difficult to study the  property pertinent to the type-II Dirac cone experimentally. Because of the existence of type-I Dirac cones at the Fermi level in graphene and its related materials\cite{Malko-2012, Wang-2016, Cheng-2017}, we speculate that these graphene-based compounds plus the simplest metallic element Li  may generate the possible system to have the type-II Dirac cone near the Fermi level with relevant electronic structures.

In this work, using the first-principles method, various structures of graphene-based compounds with Li atoms were examined. We found that the band structure of the carbon-lithium compound(C$_{4}$Li) (see Fig. 1) has two crossings; one located at  0.11 eV above the chemical potential corresponding to a type-II Dirac cone, and the other crossing is located at 0.34 eV above the chemical potential (see Fig.2). The latter crossing corresponds to a type-I Dirac cone at point W forming a node-line from W to T protected by the nonsymmorphic symmetry of the compound.  If all the Li atoms are removed from this compound, the remaining carbon atoms would have the orthorhombic structure of   bco-C$_{16}$\cite{Wang-2016}. The formation energy, phonon dispersion, Fermi surface and  the corresponding density of states (DOS) are all calculated. The  formation energy of C$_{4}$Li is smaller than that  of C$_{2}$Li$_{2}$ structure indicating that  C$_{4}$Li is thermodynamically stable. The phonon dispersion shows no imaginary frequency, and this implies that the structure is also dynamically stable. The total DOS is mainly contributed from the bands crossing the Fermi level.  The carbon atoms have more contribution to DOS while the lithium atoms have less contribution. Our x-rays diffraction (XRD) clearly show a peak characteristic for three-dimensional carbon-based structure and new peaks are found only in  C$_{4}$Li.

  \begin{figure}
  \includegraphics[width=3in]{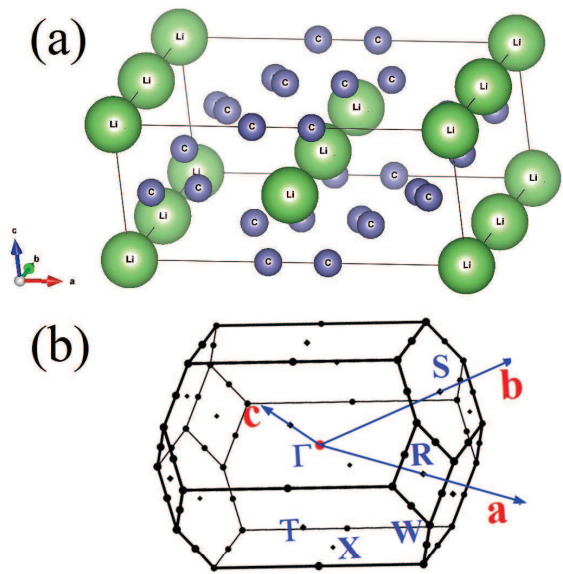}
  \caption{\label{fig:1}(colour online) (a) Crystal structure of the Carbon-Lithium C$_{4}$Li. The carbon atoms form a tree-dimensional chain with honeycomb structure when viewed from (001) plane while the lithium atoms form a chain in the (010) plane, and (b) is the corresponding primitive Brillouin zone.}
  \end{figure}

  \emph{Computational details, crystal structure and formation energy}: The  calculation of phonon frequencies is  performed by using  density functional perturbation theory (DFPT) which is implemented in quantum espresso package\cite{Giannozzi} using the pseudopotential of the  Perdew-Burke-Ernzerhof (PBE) type  with GGA exchange-correlation potential. The electronic structure calculation was performed by using  VASP (Vienna $ab$ $initio$ simulation package)\cite{Kresse-1996}. The input crystal parameters  are the following: a=7.84262 \AA,  b=4.8872 \AA, and c= 3.8680 \AA. The self-consistent calculation for the electronic  structure is  performed in the   8$\times$8$\times$8 k-grid, and the cut-off energy for the wave function is 380 Ry. The crystal structure and the XRD simulations were obtained by using the VESTA package\cite{Momma-2011}.

   \begin{table}[ht]
   \caption{Calculated parameters of different  Carbon-Lithium structures by GGA method:  The  volume of the structure per atoms, bond distance d$_{C-C}$  and d$_{Li-Li}$, and the  total formation energy per atom of C$_{2}$Li$_{2}(Immm)$ \cite{Ruschewitz-1999}, C$_{2}$Li$_{2}(Fm-3m)$\cite{Ruschewitz-1999}, C$_{4}$Li$(Imma)$, C$_{6}$Li$(P6/mmm)$\cite{Gianni-2012},C$_{12}$Li$(P6/mmm)$\cite{Avdeev-1996} and  C$_{18}$Li$(P6/mmm)$\cite{Holzwarth-1983} }
   \centering
   \begin{tabular}{c c c c c}
      \hline\hline
    Structure & Volume  & d$_{C-C}$  & d$_{Li-Li}$  & E$_{total}$  \\ [0.5ex]
       & ({\AA}$^{3}$/atom) & (\AA) &  (\AA)  & (Ry/atom) \\ [0.5ex]

   \hline\hline
   C$_{2}$Li$_{2}(Immm)$ & 7.160   & 1.2651   & 2.6647    &-0.334682 \\
   C$_{2}$Li$_{2}(Fm-3m)$ & 17.686   & 4.2179   & 2.9825    &-0.169551 \\
   C$_{4}$Li$(Imma)$       & 7.864   & 1.4575   & 2.4460    &-0.469653 \\
   C$_{6}$Li$(P6/mmm)$       & 8.497   & 1.444    & 4.3426    &-0.550839 \\
   C$_{12}$Li$(P6/mmm)$      & 8.655   & 1.442    & 4.3259    &-0.595709 \\ [1ex]
   C$_{18}$Li$(P6/mmm)$      & 8.870   & 1.432    & 4.2971    &-0.612133 \\
   \hline\hline
   \end{tabular}
   \label{table:nonlin}
   \end{table}

  The C$_{4}$Li compound crystallizes in the orthorhombic  structure with space group Imma(74). This space group is characterized by having both nonsymmorphic and the inversion symmetry. The crystal structure of C$_{4}$Li  is illustrated in Fig.\ref{fig:1}. The structure is composed of $16$ carbon atoms  and $4$ lithium atoms in the conventional unit cell.  The carbon atoms form a graphene-like  structure when viewed from (001) plane (the plane perpendicular to c axis), and the lithium atoms are located at the center of lattice  making this structure a body-center-orthorhombic structure. The nearest-neighbor bond distance of the Li-C is 2.0214 {\AA},  the bonding distance of lithium atoms is 2.4460 {\AA}, and the carbon bonding distance is 1.4575 {\AA}. Table I shows the volume of the structure per atom, the bond distances (d$_{C-C}$  and d$_{Li-Li}$) and the corresponding formation energy of different Carbon-Lithium structures which were synthesized and compared with the new structure C$_{4}$Li. The bonding distance of  \emph{C-C}  and \emph{Li-Li} for C$_{4}$Li are within the limits of carbon-lithium structures. The total energy is -188.641 Ry and the formation energy per atom is negative (-0.484573 Ry) which suggest the thermal stability of the structure. Furthermore, the formation energy(FE) of  C$_{4}$Li is  less than the FE of the known structure, the  C$_{2}$Li$_{2}(Immm)$\cite{Ruschewitz-1999} and  C$_{2}$Li$_{2}(Fm-3m)$\cite{Ruschewitz-1999} respectively, also indicating that C$_{4}$Li is thermodynamically stable.  The angle formed by the Li-C-Li is 139.049$^{o}$, and the minimal angle formed by the C-Li-C is 40.951$^{0}$.

  \begin{figure}
  \includegraphics[width=3in]{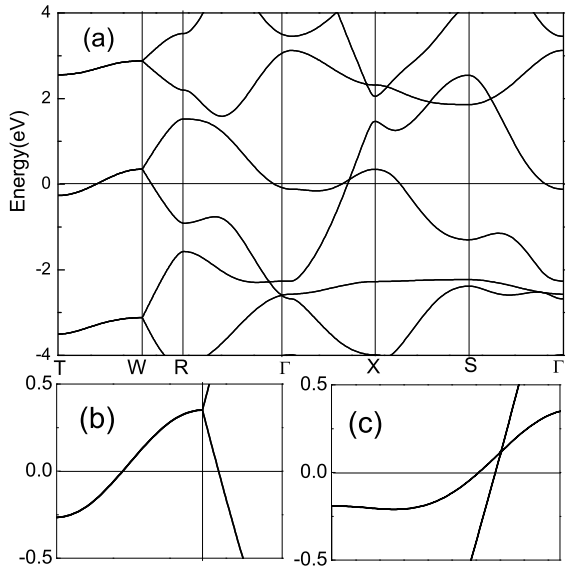}
  \caption{\label{fig:2}(colour online) (a) Band structure of C$_{4}$Li along high symmetry lines. (b) The line-node if formed from T to W located at the intersection of k$_{x}$=$\pi$ and k$_{y}$=$\pi$ mirror planes, and (c) the other crossing from $\Gamma$ to X  is the type-II Dirac cone.}
  \end{figure}

 \emph{Results  and  discussion}: The electronic structures are calculated with spin-orbit coupling interaction(SOC). Since the structure have the inversion symmetry and time-reversal symmetry the bands structure should be degenerate everywhere in the BZ.  The carbon and lithium both have week SOC, therefore, there is no significant change of the band structure with the inclusion of SOC. The band structure of C$_{4}$Li   calculated along the high symmetry lines in the first BZ is presented in Fig.\ref{fig:2}(a).  The  higher dispersive bands cross the Fermi level forming two different types of Dirac fermions.  There is  one  crossing  located at  W(0.75,-0.25,-0.25) high symmetry point and the other crossing is  located between $\Gamma$ to X(0.5,-0.5,-0.5) points in the first BZ.
  The first one shown in Fig.\ref{fig:2}(b) is originated from the crossing of the  twofold-degenerate  bands located from W to R and leads to formation of fourfold-degenerate  band  from  T to W high symmetry line respectively which is different from other known Dirac crossing because it is located at 0.346 eV above the Fermi level. The bands located from  W to T makes  a line-node at the boundary of the Brillouin zone which coincides  with the  crossing  of the mirror planes  k$_{x}$=$\pi$ and k$_{y}$=$\pi$ in the conventional cell.  This can be easily seen in Fig.4 and the symmetry analysis at higher symmetry points of the primitive cell discussed in the Supplementary Material\cite{Cuamba-2017}.

    \begin{figure}
  \includegraphics[width=3in]{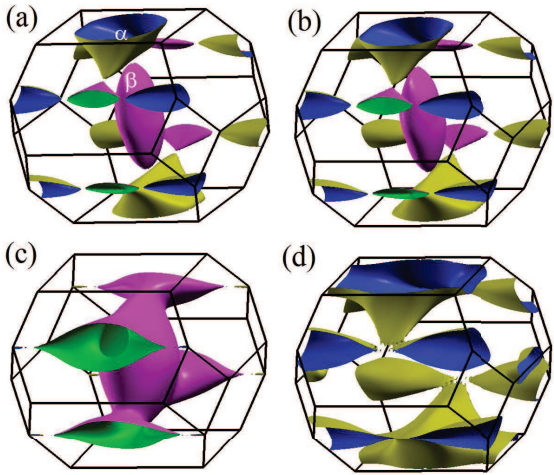}
  \caption{\label{fig:3}(colour online) Fermi surfaces of  C$_{4}$Li obtained at different energy levels. The Fermi surface (a), (b), (c) and (d) are plotted at 0 eV, 0.11 eV, 0.40 eV and -0.19 eV, respectively.}
  \end{figure}

The second crossing is the so called type-II Dirac cone  which is located at 0.11 eV above the Fermi level as illustrated in Fig.\ref{fig:2}(c).  The principal characteristic of this type of crossing appears to violate  the Lorentz invariance, which is an important principle in Higher Energy Physics. The presence  of the type-II Dirac semimetal  suggests the existence of  new physics that has never been explored before.  The band structure of  C$_{4}$Li compound  is the first  carbon-based material  to exhibit this type of electronic structure which is different from the type-I Dirac cone observed in graphene and in Na$_{3}$Bi\cite{Liu-2014-Na3Bi} and Cd$_{3}$As$_{2}$\cite{Liu-2014-Cd3As3}. The Dirac cone is titled along $\Gamma$ to X(0.5,-0.5,-0.5) high symmetry line and is  anisotropic; the dispersion has different slopes  in all direction. Along the principal diagonal of the unit cell, the anisotropy is enhanced, and this corresponds to the higher separation distance between the lithium atoms. The crossing  is located at the special  coordinate (0.3333,-0.3333, -0.3333) along the $\Gamma$ to X where the hole-like and the electron-like pockets touch each other.

 \begin{figure}
  \includegraphics[width=3in]{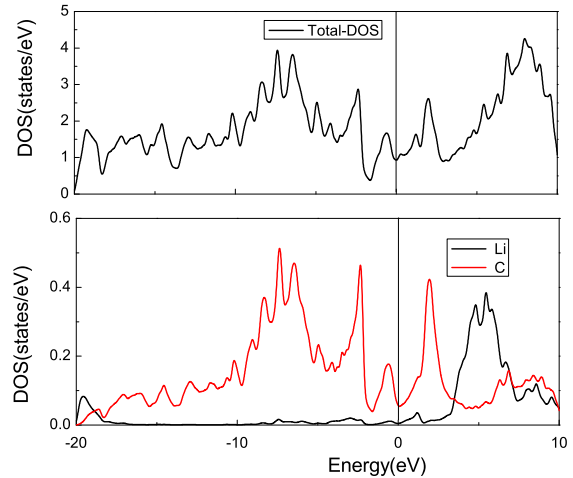}
  \caption{\label{fig:4}(colour online) (a) The total electronic density of states of C$_{4}$Li compound and (b) the partial density of states of carbon and lithium atoms.}
  \end{figure}

 The Fermi surface of C$_{4}$Li at different doping or chemical potential  are  shown in Fig.\ref{fig:3}. Fig.\ref{fig:3}(a)  illustrates the Fermi surface at zero doping, which is composed of the electron pocket $\beta$, the hole pocket $\alpha$  constituting the characteristics of  the type-II Dirac cone along the $\Gamma$ to X line, plus the other Fermi sheets corresponding to the band crossing from T-W-R high symmetry line. The two pockets $\beta$ and $\alpha$ would touch each other when the chemical potential crosses the Dirac node at the 0.11 eV(Fig.\ref{fig:3}(b)) if the compound is properly doped with electrons. The Fermi surfaces (c) and (d) in Fig.\ref{fig:3} correspond to higher electron and hole-doping respectively. With heavy electron (hole) doping, there is the disappearance of the hole (electron) pocket.

  \begin{figure}
  \includegraphics[width=3in]{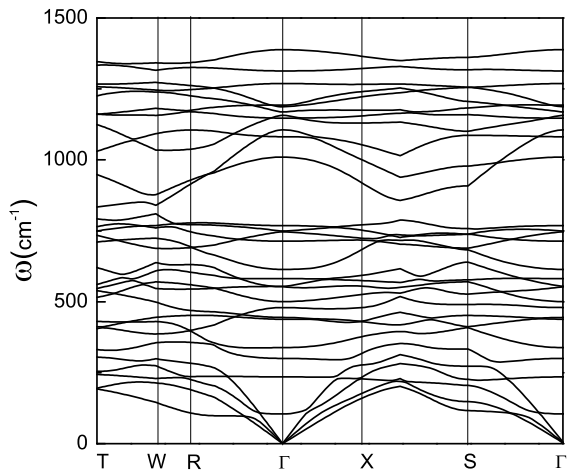}
  \caption{\label{fig:5}(colour online) Phonon dispersion for C$_{4}$Li structure.}
  \end{figure}

 The corresponding total and the partial DOS are presented in the Fig.\ref{fig:4}. The total DOS is enhanced at the Fermi level which is due to the nontrivial bands crossing the Fermi level. The carbon atoms have dominant contribution to the DOS,   while the lithium atoms have less contribution. The Dirac cones in this structure should  be easier studied  by ARPES  or magnetoresistance  experiments because all the band close to Fermi level are nontrivial, so the signal  coming from the density of states will not have any interference. The phonon dispersion of the C$_{4}$Li is presented Fig.\ref{fig:5}. The spectra do not show any imaginary frequencies, indicating that the structure is dynamically stable.

 \begin{figure}
  \includegraphics[width=3in]{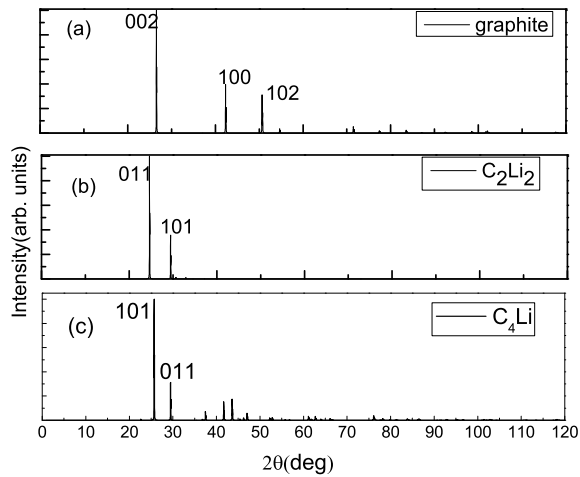}
  \caption{\label{fig:6}(colour online) X-ray diffraction pattern (XRD)  patterns of (a) graphite, (b) C$_{2}$Li$_{2}(Immm)$, and  (c) C$_{4}$Li taken at the wavelength of 1.54423 \AA.}
  \end{figure}
				
In order to compare our theory with future experiments for   the identification of   C$_{4}$Li compound, we perform X-ray diffraction simulation\cite{Momma-2011} pattern(XRD)  at  a wavelength of 1.54423 {\AA} and compare it with the known structure of Lithium-carbide, and the results are shown in  Fig.\ref{fig:6}.  The peak of C$_{4}$Li at 2$\theta$=25.69$^{0}$ is also present in graphite and in  C$_{2}$Li$_{2}(Immm)$ structure, while, the one at  29.60$^{0}$ is only present in both  C$_{4}$Li and C$_{2}$Li$_{2}(Immm)$ structures. This indicates the existence of similar characteristic of carbon-lithium structure. There are new peaks above 30$^{0}$ at 41.67$^{0}$ and 43.60$^{0}$ which are characteristics of C$_{4}$Li and the experiments could look for these peaks in order to identify this compound.

  \emph{Conclusion}:  We have predicted the new stable phase of the carbon-lithium  (C$_{4}$Li) by means of the first-principles calculation method. The obtained band structure shows two crossings, one located at W high symmetry point and the other one is the type-II Dirac cone located between the $\Gamma$ to X(0.5,-0.5,-0.5) points . The band crossing remains gapless with the inclusion of SOC due to small SOC strength of both carbon and lithium atom, and a line-node is formed at the intersection of  the mirror planes at k$_{x}$=$\pi$  and k$_{y}$=$\pi$.  Its  formation energy is shown to be lower  than that  of the known structure of C$_{2}$Li$_{2}$  indicating  the structure of C$_{4}$Li is thermodynamically stable. Our phonon dispersion calculation also indicates the compound to be dynamically stable. Hopefully, the experimentalists will be able to synthesize this compound in the future, and to discover the first material that would have the type-II Dirac cone  near the Fermi energy.   We also expect that C$_{4}$Li compound is going to exhibit exceptional topological property and to have the potential for electronic applications.

  \begin{acknowledgments}

  This work is supported by the Texas Center for Superconductivity at the University of Houston and the Robert A. Welch Foundation (Grant No. E-1146), the National Natural Science Foundation of China (Grant No. 11574108 ), the Natural Science Foundation of Anhui Province in China (Grant No. 1408085QA12), and the Natural Science Research Project of Higher Education Institutions of Anhui Province in China (Grant No. KJ2015A120).  The numerical calculations were performed at the Center for Advanced Computing and Data at the University of Houston.

  \end{acknowledgments}

\end{document}